\documentclass[12pt]{article}

\usepackage[top=35mm, bottom=25mm, left=20mm, right=20mm]{geometry}
\usepackage{url, graphicx, color, varioref, amsfonts, longtable, float}

\newcommand{\qed}{\nobreak \ifvmode \relax \else
      \ifdim\lastskip<1.5em \hskip-\lastskip
      \hskip1.5em plus0em minus0.5em \fi \nobreak
      \vrule height0.75em width0.5em depth0.25em\fi}

\title{Strength and Weakness in Grover's Quantum Search Algorithm}

\author{Ahmed Younes\footnote {ayounes2@yahoo.com}\\
Department of Mathematics and Computer Science\\
Faculty of Science\\
Alexandria University\\
Alexandria, Egypt}

\begin{document}
\maketitle
\begin{abstract}
Grover's quantum search algorithm is considered as one of the milestone in the field of quantum computing. 
The algorithm can search for a single match in a database with $N$ records  
 in $O(\sqrt{N})$ assuming that the item must exist in the database with quadratic speedup over the best known 
classical algorithm. This review paper discusses the performance of Grover's algorithm in case of multiple matches
 where the problem is expected to be easier. Unfortunately, we will find that the algorithm will fail for 
 $M>3N/4$, where $M$ is the number of matches in the list.
\end{abstract}

\section{Introduction} 
In 1996, Lov Grover \cite{grover96} presented an algorithm for searching an unstructured list of $N$ items 
with quadratic speed-up over classical algorithms. His original algorithm targets the case where a 
single match exists within the search space. Much research effort has gone into analysing and 
generalising his algorithm for multiple matches \cite{boyer96,Brassard00,Chen00b,Chen00a,Chen99}. 


This paper will review the work done by others on solving the unstructured 
search problem on quantum computers as follows: Section 2 provides the general definition of 
the unstructured search problem and some of its applications. Section 3 briefly summarises the work done so far 
in designing algorithms concerning this problem on quantum computers.
Section 4 presents Grover's algorithm in some detail and the work done by others related to his algorithm, 
analysing its performance and behaviour over the range $1\le M\le N$ for both known and unknown number of matches $M$. 
The paper ends up with a general 
conclusion in Section 5 about Grover's algorithm.

\section{Unstructured Search Problem}

Consider an unstructured list $L$ of $N$ items. For simplicity and without loss of generality we will assume that 
$N = 2^n$ for some positive integer $n$. Suppose the items in the list are labelled with the integers $\{0,1,...,N - 1\}$,  
and consider a function (oracle) $f$ which maps an item $i \in L$ to 
either 0 or 1 according to some properties this item should satisfy, i.e. $f:L \to \{ 0,1\}$. 
The problem is to find any $i \in L$ such that $f(i) = 1$ assuming that such $i$ exists in the list. 
In conventional computers, solving this problem needs $O\left({N}/{M}\right)$ calls to the oracle (query),
where $M$ is the number of items that satisfy the oracle. 

The unstructured search problem can be considered as a general domain for a wide range of applications in computer 
science, for example: 
\begin{itemize}
\item{The \it database searching problem}, where we are looking for an item in an unsorted list.
\item {The \it Boolean satisfiability problem}, where we have a Boolean expression 
with $n$ Boolean variables and we are looking for any variable assignment that satisfies this expression. 


\end{itemize}


\section{Unstructured Search on Quantum Computers}

Grover's original algorithm exploits quantum parallelism by preparing a 
{\it uniform} superposition that represents all the items 
in the list then iterates both an oracle that marks the desired item by applying a phase shift of -1 on that item 
($e^{\underline i \theta_1}$, with $\theta_1= \pi$) and nothing on the other items ($e^{\underline i \theta_2}$, with $\theta_2= 0$)  
and an operator that performs inversion about the mean (diffusion operator) 
to amplify the amplitude of the match. The process of this operator includes the operation $\left( {2\left| 0 \right\rangle \left\langle 0 \right| - I} \right)$ 
which applies a phase shift of -1 on the states 
within the superposition ($e^{\underline i \phi_1}$, with $\phi_1= \pi$) except the state 
$\left|0\right\rangle^{\otimes n}$ where it applies nothing 
($e^{\underline i \phi_2}$, with $\phi_2= 0$) (Fig. \ref{STintfig2b}) \cite{nc00a}. To maintain consistency with literature, 
this operation can also be 
written as $\left( {I - 2\left| 0 \right\rangle \left\langle 0 \right|} \right)$ which applies a phase shift of -1 on the state 
$\left|0\right\rangle^{\otimes n}$ ($e^{\underline i \phi_2}$, with $\phi_2= \pi$) and nothing on the other states of the 
superposition ($e^{\underline i \phi_1}$, with $\phi_1= 0$) {\it together with} a global phase shift of -1 (Fig. \ref{STintfig2}) \cite{Jozsa99}.

It was shown that the required number of iterations is approximately ${\pi}/{4} \sqrt {N}$ which is proved 
to be optimal to get the highest probability with the minimum number of iterations \cite{Zalka99}, 
if there is exactly one match in the search space.

In \cite{BK02,Gal00,Grover98a,Jozsa99,long01}, Grover's algorithm is generalised by showing that the uniform 
superposition can be replaced by almost any arbitrary superposition and the phase shifts applied 
by the oracle and the diffusion operator ($e^{\underline i \theta_1},e^{\underline i\theta_2},e^{\underline i \phi_1}$ and $e^{\underline i \phi_2}$) can be generalised to deal with 
the arbitrary superposition and/or to increase the probability of success even with a factor increase in the number of iterations 
to still run in $O(\sqrt{N/M})$.
These give a larger class of algorithms for amplitude amplification using variable operators 
from which Grover's algorithm was shown to be a special case.


In another research direction, work has been done trying to generalise Grover's algorithm with a uniform superposition 
for the case where there are a known number of multiple matches in the search space 
\cite{boyer96,Chen00a,Chen99}, 
where it was shown that the required number of iterations is approximately 
${\pi}/{4}\sqrt {{N}/{M}}$ for small ${M}/{N}$. The required number of iterations will increase for $M>{N}/{2}$, 
i.e. the problem will be harder where it might be expected to be easier \cite{nc00a}. Other work has been done 
for a known number of multiple matches with arbitrary superposition and phase shifts \cite{Biron98,Brassard00,hoyer00,Li01,Mosca98} where the same problem 
for multiple matches occurs. In \cite{Brassard00,Brassard98,Mosca98}, a hybrid algorithm was presented to deal with this problem. 
It applies Grover's fixed operators algorithm for ${\pi}/{4}\sqrt {{N}/{M}}$ times then applies one more iteration using different oracle 
and diffusion operator by replacing the standard phase shifts $\theta_1=\pi, \phi_1=\pi$ with accurately calculated 
phase shifts $\theta_1^{'}$ and $\phi_1^{'}$ according to the knowledge of the number of matches $M$ to get the solution with probability 
close to certainty. Using this algorithm will increase the hardware cost since we have to build one more oracle and one more  
diffusion operator for each particular $M$. For the sake of practicality, the operators should be fixed for any 
given $M$ and are able to handle the problem with high probability whether or not $M$ is known in advance.

In case of multiple matches, where the number of matches is unknown, an algorithm for estimating the number of matches 
(known as {\it quantum counting algorithm}) was presented \cite{Brassard98,Mosca98}. In \cite{boyer96}, 
another algorithm was presented to find a match even if the number of matches is unknown which will be able to work 
if $M$ lies within the range $1\le M \le 3N/4$, otherwise it was suggested to use standard sampling techniques.
 
Many Grover-like algorithms for solving a wide range of applications have been presented. For example, 
an algorithm for the scheduling problem (intersection problem) \cite{grover02} and an algorithm for minimum finding \cite{Durr96}.


\section{Grover's Quantum Search Algorithm}

\subsection{Number of Matches is Known}

In this section, we will present Grover's algorithm for searching a list of size $N$ with $M$ matches 
such that $1 \le M \le N$. We assume that $M$ is known in advance. For our purposes, the analysis will 
concentrate on the behaviour of the algorithm if iterated once, then the behaviour after $q_G$ iterations. 

\subsubsection{Iterating the Algorithm Once}

\begin{center}
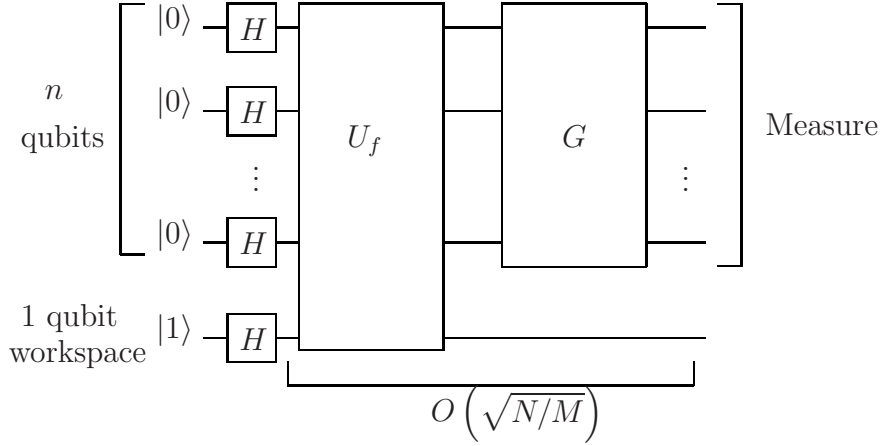
\begin{figure} [t]
\begin{center}
\setlength{\unitlength}{3947sp}%
\begingroup\makeatletter\ifx\SetFigFont\undefined%
\gdef\SetFigFont#1#2#3#4#5{%
  \reset@font\fontsize{#1}{#2pt}%
  \fontfamily{#3}\fontseries{#4}\fontshape{#5}%
  \selectfont}%
\fi\endgroup%
\begin{picture}(4650,2691)(2551,-2890)
\thinlines
{\color[rgb]{0,0,0}\put(3676,-361){\line( 1, 0){150}}
}%
{\color[rgb]{0,0,0}\put(3676,-886){\line( 1, 0){150}}
}%
{\color[rgb]{0,0,0}\put(3676,-1711){\line( 1, 0){150}}
}%
{\color[rgb]{0,0,0}\put(3151,-211){\line( 0,-1){1575}}
}%
{\color[rgb]{0,0,0}\put(3151,-211){\line( 1, 0){150}}
}%
{\color[rgb]{0,0,0}\put(3151,-1786){\line( 1, 0){150}}
}%
{\color[rgb]{0,0,0}\put(3826,-511){\framebox(300,300){}}
}%
{\color[rgb]{0,0,0}\put(3826,-1036){\framebox(300,300){}}
}%
{\color[rgb]{0,0,0}\put(3826,-1861){\framebox(300,300){}}
}%
{\color[rgb]{0,0,0}\put(4126,-361){\line( 1, 0){150}}
}%
{\color[rgb]{0,0,0}\put(4126,-886){\line( 1, 0){150}}
}%
{\color[rgb]{0,0,0}\put(4126,-1711){\line( 1, 0){150}}
}%
{\color[rgb]{0,0,0}\put(4276,-2386){\framebox(900,2175){}}
}%
{\color[rgb]{0,0,0}\put(5176,-361){\line( 1, 0){375}}
}%
{\color[rgb]{0,0,0}\put(5176,-886){\line( 1, 0){375}}
}%
{\color[rgb]{0,0,0}\put(5176,-1711){\line( 1, 0){375}}
}%
{\color[rgb]{0,0,0}\put(5551,-1861){\framebox(900,1650){}}
}%
{\color[rgb]{0,0,0}\put(6451,-361){\line( 1, 0){375}}
}%
{\color[rgb]{0,0,0}\put(6451,-886){\line( 1, 0){375}}
}%
{\color[rgb]{0,0,0}\put(6451,-1711){\line( 1, 0){375}}
}%
{\color[rgb]{0,0,0}\put(5176,-2311){\line( 1, 0){1650}}
}%
{\color[rgb]{0,0,0}\put(6901,-211){\line( 1, 0){150}}
}%
{\color[rgb]{0,0,0}\put(7051,-211){\line( 0,-1){1650}}
}%
{\color[rgb]{0,0,0}\put(6901,-1861){\line( 1, 0){150}}
}%
{\color[rgb]{0,0,0}\put(3826,-2461){\framebox(300,300){}}
}%
{\color[rgb]{0,0,0}\put(3676,-2311){\line( 1, 0){150}}
}%
{\color[rgb]{0,0,0}\put(4126,-2311){\line( 1, 0){150}}
}%
{\color[rgb]{0,0,0}\put(4201,-2611){\line( 1, 0){2550}}
}%
{\color[rgb]{0,0,0}\put(4201,-2461){\line( 0,-1){150}}
}%
{\color[rgb]{0,0,0}\put(6751,-2461){\line( 0,-1){150}}
}%

\put(3976,-1386){$\vdots$}
\put(6676,-1386){$\vdots$}

\put(3376,-361){$\left| 0 \right\rangle$}%
\put(3376,-886){$\left| 0 \right\rangle$}%
\put(3376,-1711){$\left| 0 \right\rangle$}%
\put(3376,-2311){$\left| 1 \right\rangle$}%

\put(2676,-811){$n$}%
\put(2526,-1111){qubits}%

\put(2526,-2236){1 qubit}%
\put(2451,-2461){workspace}%

\put(3901,-436){$H$}%
\put(3901,-961){$H$}%
\put(3901,-1786){$H$}%
\put(3901,-2386){$H$}%

\put(4576,-1111){$U_f$}%
\put(5926,-1111){$G$}%
\put(7201,-1036){Measure}%

\put(5101,-2836){$O\left( {\sqrt {{N}/{M}} } \right)$}

\end{picture}
\end{center}
\caption{Quantum circuit for Grover's algorithm.}
\label{SIntfig1}
\end{figure}
\end{center}

For a list of size $N=2^n$, the steps of the algorithm can be understood 
as follows (its quantum circuit is shown in Fig. \ref{SIntfig1}): 

\begin{itemize}

\item[1-]{\it Register Preparation}. Prepare a quantum register of $n+1$ 
qubits. The first $n$ qubits all in state $\left| 0 \right\rangle$ and the extra qubit in state 
$\left| 1 \right\rangle$ where it will be used as a workspace for evaluating the oracle $U_f$. The state of the 
system $\left| {W_0^{(G,1)} } \right\rangle$ can be written 
as follows, where the subscript number refers to the step within the iteration. $(G,1)$ in the superscript is 
the diffusion operator used in the algorithm which will be defined later and the iteration number respectively:

\begin{equation}
\label{SInteq1}
\left| {W_0^{(G,1)} } \right\rangle = \left| 0 \right\rangle ^{ \otimes n} \otimes 
\left| 1 \right\rangle. 
\end{equation}

\item[2-] {\it Register Initialisation}. Apply the Hadamard gate on each of the $n+1$ qubits in 
parallel so that the first $n$ qubits will contain the $2^{n}$ states representing the list and the extra qubit will be in the state 
${{\left( {\left| 0 \right\rangle  - \left| 1 \right\rangle } \right)} \mathord{\left/
 {\vphantom {{\left( {\left| 0 \right\rangle  - \left| 1 \right\rangle } \right)} {\sqrt 2 }}} \right.
 \kern-\nulldelimiterspace} {\sqrt 2 }}$, where $i$ is the integer representation of the items in the list:

\begin{equation}
\label{SInteq2}
\left| {W_1^{(G,1)} } \right\rangle = 
{H^{ \otimes n+1}}\, \left| {W_0^{(G,1)} } \right\rangle = 
{\frac{1}{\sqrt{N}}\sum\limits_{i = 0}^{N - 1} {\left| i \right\rangle }\otimes 
\left( \frac{{\left| 0 \right\rangle  - \left| 1 \right\rangle }}{\sqrt{2}} \right) }  .
\end{equation}

\item[3-] {\it Applying the Oracle and Changing Sign}. Apply the oracle $U_{f}$ 
that gives the amplitudes of the matches a phase shift of $-1$ ($e^{\underline i \pi}$), i.e. 
$U_f \left| i \right\rangle  \to \left( { - 1} \right)^{f(i)} \left| i \right\rangle$, so that, 

\begin{equation}
\label{SInteq3}
\left| {W_2^{(G,1)} } \right\rangle  = 
U_f \left| {W_1^{(G,1)} } \right\rangle  = 
\frac{1}{{\sqrt {N} }}\sum\limits_{i = 0}^{N - 1} {\left| i \right\rangle  \otimes 
\left( \frac{{\left| {0 \oplus f\left( i \right)} \right\rangle  
- \left| {1 \oplus f\left( i \right)} \right\rangle }}{\sqrt 2} \right)}.
\end{equation}

Notice that, if $f(i)=0$, then $\left| {0 \oplus f\left( i \right)} \right\rangle = \left| 0 \right\rangle$ and
$\left| {1 \oplus f\left( i \right)} \right\rangle = \left| 1 \right\rangle$, and if $f(i)=1$, 
then $\left| {0 \oplus f\left( i \right)} \right\rangle = \left| 1 \right\rangle$ and
$\left| {1 \oplus f\left( i \right)} \right\rangle = \left| 0 \right\rangle$. Assume that 
$\sum\nolimits_i {{'}} $ denotes a sum over $i$ which are desired matches 
and $\sum\nolimits_i {{''}}$ denotes a sum over $i$ which are undesired items in the list. 
So, the system  $\left| {W_2^{(G,1)} } \right\rangle$ shown in Eqn. \ref{SInteq3} can be re-written as follows:

\begin{equation}
\label{SInteq4}
\left| {W_2^{(G,1)} } \right\rangle = 
\frac{1}{{\sqrt {N} }}\sum\limits_{i = 0}^{N - 1} {''\left| i \right\rangle  
\otimes \left( \frac{{\left| 0 \right\rangle  - \left| 1 \right\rangle }}{\sqrt 2} \right)}  
- \frac{1}{{\sqrt {N} }}\sum\limits_{i = 0}^{N - 1} {'\left| i \right\rangle  
\otimes \left( \frac{{\left| 0 \right\rangle  - \left| 1 \right\rangle }}{\sqrt 2} \right)}. 
\end{equation}

Notice the change of the sign for the states that represent the matches in the search space (phase shift of -1), with no change
to the state of the extra qubit workspace, which can be removed from the system for simplicity. We end-up with 
a system as follows:

\begin{equation}
\label{SInteq5}
\left| {W_2^{(G,1)} } \right\rangle  = 
\frac{1}{{\sqrt {N} }}\sum\limits_{i = 0}^{N - 1} {''\left| i \right\rangle }  
- \frac{1}{{\sqrt {N} }}\sum\limits_{i = 0}^{N - 1} {'\left| i \right\rangle }. 
\end{equation}


\pagebreak
\item[4-]{\it Inversion about the Mean}\label{GMean}. Apply the {\it Diffusion Operator} $G$  on the first $n$ 
qubits. The diagonal representation of $G$ can take this form (its quantum circuit is as 
shown in Fig. \ref{STintfig2b} and its quantum circuit with a global phase shift factor of -1 \cite{nc00a} is as 
shown in Fig. \ref{STintfig2}):

\begin{equation}
\label{SInteqn6}
G = H^{ \otimes n }\left( {2\left| 0 \right\rangle \left\langle 0 \right| 
- I_n} \right)H^{ \otimes n},
\end{equation}

\noindent
where the vector $\left|0\right\rangle$ used in Eqn. \ref{SInteqn6} is of length $N=2^n$, and 
$I_{n}$ is the identity matrix of size $2^{n}\times 2^{n}$. 
Consider a general system $\left|\psi\right\rangle$ of $n$-qubit quantum register:

\begin{equation}
\left| \psi  \right\rangle  = \sum\limits_{j = 0}^{N - 1} {\alpha _j \left| j \right\rangle }.
\end{equation}

The effect of applying $G$ on $\left| \psi  \right\rangle$ produces,

\begin{equation}
G\left| \psi  \right\rangle  = \sum\limits_{j = 0}^{N - 1} {\left[ { - \alpha _j  + 2\left\langle \alpha  \right\rangle } \right]\left| j \right\rangle },
\end{equation}
where, $\left\langle \alpha \right\rangle = \frac{1}{N}\sum\nolimits_{j = 0}^{N-1} {\alpha _j }$ 
is the mean of the amplitudes of the states in the superposition, 
i.e. each amplitude $\alpha _j $ will be transformed according to the following relation:

\begin{equation}
\label{SInteqn7}
\alpha _j \to \left[ { - \alpha _j + 2\left\langle \alpha \right\rangle } 
\right].
\end{equation}

From Eqn. \ref{SInteq5} we can see that there are $M$ states with amplitude 
${-1}/{\sqrt{N}}$ and $N-M$ states with amplitude 
${1}/{\sqrt{N}}$, so the mean $\left\langle \alpha  \right\rangle$ can be calculated as follows:

\begin{equation}
\label{SInteqn8}
\begin{array}{l}

\left\langle \alpha \right\rangle = \frac{1}{N}\left( 
{M\left( {\frac{ - 1}{\sqrt N }} \right) + (N - M)\left( {\frac{1}{\sqrt N 
}} \right)} \right)\\
 \,\,\,\,\,\,\,\,\,\,= \frac{1}{{\sqrt N }}\left( {1 - \frac{{2M}}{N}} \right).
 \end{array}
\end{equation}

\begin{center}
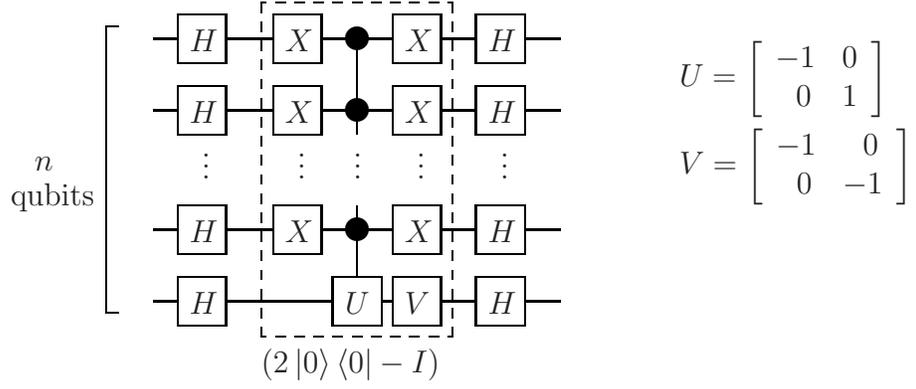
\begin{figure}  [t]
\begin{center}

\setlength{\unitlength}{3947sp}%
\begingroup\makeatletter\ifx\SetFigFont\undefined%
\gdef\SetFigFont#1#2#3#4#5{%
  \reset@font\fontsize{#1}{#2pt}%
  \fontfamily{#3}\fontseries{#4}\fontshape{#5}%
  \selectfont}%
\fi\endgroup%
\begin{picture}(4200,2337)(3226,-2461)
{\color[rgb]{0,0,0}\thinlines
\put(5401,-361){\circle*{150}}
}%
{\color[rgb]{0,0,0}\put(5401,-811){\circle*{150}}
}%
{\color[rgb]{0,0,0}\put(5401,-1561){\circle*{150}}
}%
{\color[rgb]{0,0,0}\put(3826,-286){\line( 0,-1){1800}}
}%
{\color[rgb]{0,0,0}\put(3826,-286){\line( 1, 0){ 75}}
}%
{\color[rgb]{0,0,0}\put(3826,-2086){\line( 1, 0){ 75}}
}%
{\color[rgb]{0,0,0}\put(5401,-361){\line( 0,-1){600}}
}%
{\color[rgb]{0,0,0}\put(5401,-1411){\line( 0,-1){450}}
}%
{\color[rgb]{0,0,0}\put(5251,-2161){\framebox(300,300){}}
}%
{\color[rgb]{0,0,0}\put(4876,-1711){\framebox(300,300){}}
}%
{\color[rgb]{0,0,0}\put(4876,-961){\framebox(300,300){}}
}%
{\color[rgb]{0,0,0}\put(4876,-511){\framebox(300,300){}}
}%
{\color[rgb]{0,0,0}\put(5626,-1711){\framebox(300,300){}}
}%
{\color[rgb]{0,0,0}\put(5626,-961){\framebox(300,300){}}
}%
{\color[rgb]{0,0,0}\put(5626,-511){\framebox(300,300){}}
}%
{\color[rgb]{0,0,0}\put(5626,-2161){\framebox(300,300){}}
}%
{\color[rgb]{0,0,0}\put(4276,-1711){\framebox(300,300){}}
}%
{\color[rgb]{0,0,0}\put(4276,-2161){\framebox(300,300){}}
}%
{\color[rgb]{0,0,0}\put(4276,-961){\framebox(300,300){}}
}%
{\color[rgb]{0,0,0}\put(4276,-511){\framebox(300,300){}}
}%
{\color[rgb]{0,0,0}\put(6151,-511){\framebox(300,300){}}
}%
{\color[rgb]{0,0,0}\put(6151,-961){\framebox(300,300){}}
}%
{\color[rgb]{0,0,0}\put(6151,-1711){\framebox(300,300){}}
}%
{\color[rgb]{0,0,0}\put(6151,-2161){\framebox(300,300){}}
}%

\put(4426,-1236){$\vdots$}
\put(5026,-1236){$\vdots$}
\put(5380,-1236){$\vdots$}
\put(5776,-1236){$\vdots$}
\put(6301,-1236){$\vdots$}


{\color[rgb]{0,0,0}\put(4126,-361){\line( 1, 0){150}}
}%
{\color[rgb]{0,0,0}\put(4576,-361){\line( 1, 0){300}}
}%
{\color[rgb]{0,0,0}\put(5176,-361){\line( 1, 0){450}}
}%
{\color[rgb]{0,0,0}\put(5926,-361){\line( 1, 0){225}}
}%
{\color[rgb]{0,0,0}\put(6451,-361){\line( 1, 0){225}}
}%
{\color[rgb]{0,0,0}\put(4576,-811){\line( 1, 0){300}}
}%
{\color[rgb]{0,0,0}\put(4576,-1561){\line( 1, 0){300}}
}%
{\color[rgb]{0,0,0}\put(5176,-811){\line( 1, 0){450}}
}%
{\color[rgb]{0,0,0}\put(5926,-811){\line( 1, 0){225}}
}%
{\color[rgb]{0,0,0}\put(6451,-811){\line( 1, 0){225}}
}%
{\color[rgb]{0,0,0}\put(4126,-811){\line( 1, 0){150}}
}%
{\color[rgb]{0,0,0}\put(5176,-1561){\line( 1, 0){450}}
}%
{\color[rgb]{0,0,0}\put(5926,-1561){\line( 1, 0){225}}
}%
{\color[rgb]{0,0,0}\put(6451,-1561){\line( 1, 0){225}}
}%
{\color[rgb]{0,0,0}\put(4126,-1561){\line( 1, 0){150}}
}%
{\color[rgb]{0,0,0}\put(4126,-2011){\line( 1, 0){150}}
}%
{\color[rgb]{0,0,0}\put(4576,-2011){\line( 1, 0){675}}
}%
{\color[rgb]{0,0,0}\put(5551,-2011){\line( 1, 0){ 75}}
}%
{\color[rgb]{0,0,0}\put(5926,-2011){\line( 1, 0){225}}
}%
{\color[rgb]{0,0,0}\put(6451,-2011){\line( 1, 0){225}}
}%
{\color[rgb]{0,0,0}\put(4801,-2236){\dashbox{60}(1200,2100){}}
}%
\put(3226,-1411){qubits}%
\put(3376,-1186){$n$}%
\put(5326,-2086){$U$}%
\put(7426,-661){$U = \left[ {\begin{array}{*{20}c}
   { - 1} & 0  \\
   {\,\,\,0} & 1  \\
\end{array}} \right]$}%
\put(7426,-1211){$V = \left[ {\begin{array}{*{20}c}
   { - 1} & {\,\,\,0}  \\
   {\,\,\,0} & {-1}  \\
\end{array}} \right]$}%

\put(5701,-436){$X$}%
\put(5701,-886){$X$}%
\put(5701,-1636){$X$}%
\put(5701,-2086){$V$}%
\put(4951,-1636){$X$}%
\put(4951,-886){$X$}%
\put(4951,-436){$X$}%
\put(4351,-1636){$H$}%
\put(4351,-2086){$H$}%
\put(4351,-886){$H$}%
\put(4351,-436){$H$}%
\put(6226,-2086){$H$}%
\put(6226,-1636){$H$}%
\put(6226,-886){$H$}%
\put(6226,-436){$H$}%
\put(4800,-2461){$\left( {2\left| 0 \right\rangle \left\langle 0 \right| - I} \right)$}%

\end{picture}

\end{center}
\caption{Quantum circuit for the diffusion operator $G$ over $n$ qubits.}
\label{STintfig2b}
\end{figure}
\end{center}

\begin{center}
\begin{figure}  [t]
\begin{center}
\setlength{\unitlength}{3947sp}%
\begingroup\makeatletter\ifx\SetFigFont\undefined%
\gdef\SetFigFont#1#2#3#4#5{%
  \reset@font\fontsize{#1}{#2pt}%
  \fontfamily{#3}\fontseries{#4}\fontshape{#5}%
  \selectfont}%
\fi\endgroup%
\begin{picture}(4200,2337)(3226,-2461)
{\color[rgb]{0,0,0}\thinlines
\put(5401,-361){\circle*{150}}
}%
{\color[rgb]{0,0,0}\put(5401,-811){\circle*{150}}
}%
{\color[rgb]{0,0,0}\put(5401,-1561){\circle*{150}}
}%
{\color[rgb]{0,0,0}\put(5401,-2011){\circle{150}}
}%
{\color[rgb]{0,0,0}\put(3976,-511){\framebox(300,300){}}
}%
{\color[rgb]{0,0,0}\put(4276,-361){\line( 1, 0){225}}
}%
{\color[rgb]{0,0,0}\put(4501,-511){\framebox(300,300){}}
}%
{\color[rgb]{0,0,0}\put(3976,-961){\framebox(300,300){}}
}%
{\color[rgb]{0,0,0}\put(4276,-811){\line( 1, 0){225}}
}%
{\color[rgb]{0,0,0}\put(4501,-961){\framebox(300,300){}}
}%
{\color[rgb]{0,0,0}\put(3976,-1711){\framebox(300,300){}}
}%
{\color[rgb]{0,0,0}\put(4276,-1561){\line( 1, 0){225}}
}%
{\color[rgb]{0,0,0}\put(4501,-1711){\framebox(300,300){}}
}%
{\color[rgb]{0,0,0}\put(3901,-361){\line( 1, 0){ 75}}
}%
{\color[rgb]{0,0,0}\put(3901,-811){\line( 1, 0){ 75}}
}%
{\color[rgb]{0,0,0}\put(3901,-1561){\line( 1, 0){ 75}}
}%
{\color[rgb]{0,0,0}\put(3826,-286){\line( 0,-1){1800}}
}%
{\color[rgb]{0,0,0}\put(3826,-286){\line( 1, 0){ 75}}
}%
{\color[rgb]{0,0,0}\put(4501,-2161){\framebox(300,300){}}
}%
{\color[rgb]{0,0,0}\put(3976,-2161){\framebox(300,300){}}
}%
{\color[rgb]{0,0,0}\put(4276,-2011){\line( 1, 0){225}}
}%
{\color[rgb]{0,0,0}\put(3901,-2011){\line( 1, 0){ 75}}
}%
{\color[rgb]{0,0,0}\put(3826,-2086){\line( 1, 0){ 75}}
}%
{\color[rgb]{0,0,0}\put(5401,-361){\line( 0,-1){600}}
}%
{\color[rgb]{0,0,0}\put(5401,-1411){\line( 0,-1){675}}
}%
{\color[rgb]{0,0,0}\put(4951,-2161){\framebox(300,300){}}
}%
{\color[rgb]{0,0,0}\put(5551,-2161){\framebox(300,300){}}
}%
{\color[rgb]{0,0,0}\put(6826,-2011){\line( 1, 0){ 75}}
}%
{\color[rgb]{0,0,0}\put(6826,-1561){\line( 1, 0){ 75}}
}%
{\color[rgb]{0,0,0}\put(6826,-811){\line( 1, 0){ 75}}
}%
{\color[rgb]{0,0,0}\put(6826,-361){\line( 1, 0){ 75}}
}%
{\color[rgb]{0,0,0}\put(6526,-511){\framebox(300,300){}}
}%
{\color[rgb]{0,0,0}\put(6526,-961){\framebox(300,300){}}
}%
{\color[rgb]{0,0,0}\put(6526,-1711){\framebox(300,300){}}
}%
{\color[rgb]{0,0,0}\put(6526,-2161){\framebox(300,300){}}
}%
{\color[rgb]{0,0,0}\put(6301,-2011){\line( 1, 0){225}}
}%
{\color[rgb]{0,0,0}\put(6301,-1561){\line( 1, 0){225}}
}%
{\color[rgb]{0,0,0}\put(6301,-811){\line( 1, 0){225}}
}%
{\color[rgb]{0,0,0}\put(6301,-361){\line( 1, 0){225}}
}%
{\color[rgb]{0,0,0}\put(6001,-511){\framebox(300,300){}}
}%
{\color[rgb]{0,0,0}\put(6001,-961){\framebox(300,300){}}
}%
{\color[rgb]{0,0,0}\put(6001,-1711){\framebox(300,300){}}
}%
{\color[rgb]{0,0,0}\put(6001,-2161){\framebox(300,300){}}
}%
{\color[rgb]{0,0,0}\put(4801,-361){\line( 1, 0){1200}}
}%
{\color[rgb]{0,0,0}\put(4801,-811){\line( 1, 0){1200}}
}%
{\color[rgb]{0,0,0}\put(4801,-1561){\line( 1, 0){1200}}
}%
{\color[rgb]{0,0,0}\put(4801,-2011){\line( 1, 0){150}}
}%
{\color[rgb]{0,0,0}\put(5251,-2011){\line( 1, 0){300}}
}%
{\color[rgb]{0,0,0}\put(5851,-2011){\line( 1, 0){150}}
}%

\put(4126,-1236){$\vdots$}
\put(4651,-1236){$\vdots$}
\put(5380,-1236){$\vdots$}
\put(6676,-1236){$\vdots$}
\put(6151,-1236){$\vdots$}

\put(4051,-436){$H$}%
\put(4051,-886){$H$}%
\put(4051,-1636){$H$}%
\put(4576,-436){$X$}%
\put(4576,-886){$X$}%
\put(4576,-1636){$X$}%
\put(4051,-2086){$H$}%
\put(4576,-2086){$X$}%
\put(3226,-1411){qubits}%
\put(3376,-1186){$n$}%
\put(6601,-436){$H$}%
\put(6601,-886){$H$}%
\put(6601,-1636){$H$}%
\put(6601,-2086){$H$}%
\put(6076,-886){$X$}%
\put(6076,-436){$X$}%
\put(6076,-1636){$X$}%
\put(6076,-2086){$X$}%
\put(5026,-2086){$H$}%
\put(5626,-2086){$H$}%

{\color[rgb]{0,0,0}\put(4401,-2236){\dashbox{60}(2000,2100){}}
}%

\put(4800,-2461){$ \left( {I - 2\left| 0 \right\rangle \left\langle 0 \right|} \right)$}%

\end{picture}

\end{center}
\caption{Quantum circuit for the diffusion operator $G$ over $n$ qubits with a global phase shift factor of -1 \cite{nc00a}.}
\label{STintfig2}
\end{figure}
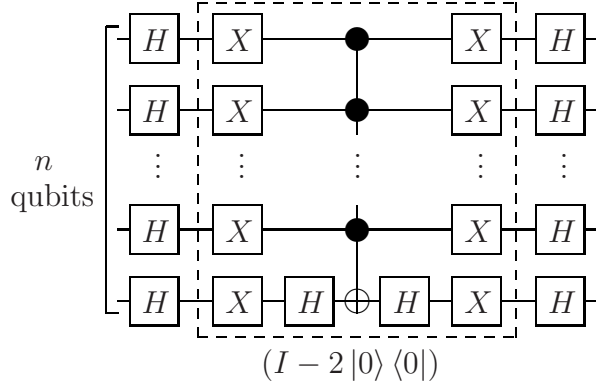
\end{center}
 
The effect of applying $G$ on the system $\left| {W_2^{(G,1)} } \right\rangle$ shown in Eqn. \ref{SInteq5} 
can be understood as follows:

\begin{itemize}

\item[a-] The $M$ negative sign amplitudes (solutions) will be transformed from 
${-1}/{\sqrt{N}}$ to $a_1^G$, where $a_1^G$ is calculated as follows: Substitute $\alpha _j = 
{ - 1}/{\sqrt N }$ and $\left\langle \alpha  \right\rangle$ 
from Eqn. \ref{SInteqn8} in Eqn. \ref{SInteqn7} we get:

\begin{equation}
\label{SInteqn9}
\begin{array}{l}
 a_1^G = - \left( {\frac{ - 1}{\sqrt N }} \right) + \frac{2}{{\sqrt N }}\left( {1 - \frac{{2M}}{N}} \right) \\ 
\,\,\,\,\,\,\,\, = \frac{1}{\sqrt N }\left( {3 - \frac{4M}{N}} \right). \\
 \end{array}
\end{equation}

\item[b-] The $(N-M)$ positive sign amplitudes will be transformed from 
${1}/{\sqrt{N}}$ to $b_1^G$, where $b_1^G$ is calculated as follows: Substitute $\alpha _j = 
{ 1}/{\sqrt N }$ and $\left\langle \alpha  \right\rangle$ 
from Eqn. \ref{SInteqn8} in Eqn. \ref{SInteqn7} we get:

\begin{equation}
\label{SInteqn10}
\begin{array}{l}
 b_1^G = - \left( {\frac{1}{\sqrt N }} \right) +\frac{2}{{\sqrt N }}\left( {1 - \frac{{2M}}{N}} \right) \\ 
 \,\,\,\,\,\,\,\, = \frac{1}{\sqrt N }\left( {1 - \frac{4M}{N}} \right). \\ 
 \end{array}
\end{equation}

\end{itemize}

The new system $\left| {W_3^{(G,1)} } \right\rangle$ after applying $G$ can be written as follows, and the mechanism of 
amplifying the amplitudes can be understood as shown in Fig. \ref{GrovMech}:

\begin{center}
\begin{figure} [t]
\begin{center}

\setlength{\unitlength}{3947sp}%
\begingroup\makeatletter\ifx\SetFigFont\undefined%
\gdef\SetFigFont#1#2#3#4#5{%
  \reset@font\fontsize{#1}{#2pt}%
  \fontfamily{#3}\fontseries{#4}\fontshape{#5}%
  \selectfont}%
\fi\endgroup%
\begin{picture}(2589,4155)(-176,-2611)
\thicklines
{\color[rgb]{0,0,0}\put(301,1064){\line( 0,-1){225}}
}%
{\color[rgb]{0,0,0}\put(601,1064){\line( 0,-1){225}}
}%
{\color[rgb]{0,0,0}\put(901,1064){\line( 0,-1){225}}
}%
{\color[rgb]{0,0,0}\put(1201,1064){\line( 0,-1){225}}
}%
{\color[rgb]{0,0,0}\put(1501,1064){\line( 0,-1){225}}
}%
{\color[rgb]{0,0,0}\put(1801,1064){\line( 0,-1){225}}
}%
{\color[rgb]{0,0,0}\put(2101,1064){\line( 0,-1){225}}
}%
{\color[rgb]{0,0,0}\put(  1,-586){\line( 0,-1){225}}
}%
{\color[rgb]{0,0,0}\put(301,-586){\line( 0,-1){225}}
}%
{\color[rgb]{0,0,0}\put(601,-586){\line( 0,-1){225}}
}%
{\color[rgb]{0,0,0}\put(901,-586){\line( 0,-1){225}}
}%
{\color[rgb]{0,0,0}\put(1201,-586){\line( 0,-1){225}}
}%
{\color[rgb]{0,0,0}\put(1501,-586){\line( 0,-1){225}}
}%
{\color[rgb]{0,0,0}\put(1801,-586){\line( 0,-1){225}}
}%
{\color[rgb]{0,0,0}\put(  1,-2466){\line( 0, 1){115}}
}%
{\color[rgb]{0,0,0}\put(301,-2466){\line( 0, 1){115}}
}%
{\color[rgb]{0,0,0}\put(601,-2466){\line( 0, 1){115}}
}%
{\color[rgb]{0,0,0}\put(901,-2466){\line( 0, 1){115}}
}%
{\color[rgb]{0,0,0}\put(1201,-2466){\line( 0, 1){115}}
}%
{\color[rgb]{0,0,0}\put(1501,-2466){\line( 0, 1){115}}
}%
{\color[rgb]{0,0,0}\put(1801,-2466){\line( 0, 1){115}}
}%
{\color[rgb]{0,0,0}\put(2101,-811){\line( 0,-1){225}}
}%
{\color[rgb]{0,0,0}\put(2101,-1806){\line( 0,-1){655}}
}%
\thinlines
{\color[rgb]{0,0,0}\put(1051,389){\vector( 0,-1){450}}
}%
{\color[rgb]{0,0,0}\put(1051,-1261){\vector( 0,-1){450}}
}%
{\color[rgb]{0,0,0}\put(2251,1364){\vector(-2,-3){150}}
}%
\thicklines
{\color[rgb]{0,0,0}\put(  1,1064){\line( 0,-1){225}}
}%
\thinlines
{\color[rgb]{0,0,0}\put(-149,839){\line( 1, 0){2550}}
}%
{\color[rgb]{0,0,0}\put(-149,-811){\line( 1, 0){2550}}
}%
{\color[rgb]{0,0,0}\put(-149,-2461){\line( 1, 0){2550}}
}%
{\color[rgb]{0,0,0}\multiput(-164,-666)(119.75610,0.00000){21}{\line( 1, 0){ 59.878}}
}%
\put(-149,689){\makebox(0,0)[lb]{\smash{\SetFigFont{10}{12.0}{\rmdefault}{\mddefault}{\updefault}{\color[rgb]{0,0,0}000}%
}}}
\put(151,689){\makebox(0,0)[lb]{\smash{\SetFigFont{10}{12.0}{\rmdefault}{\mddefault}{\updefault}{\color[rgb]{0,0,0}001}%
}}}
\put(451,689){\makebox(0,0)[lb]{\smash{\SetFigFont{10}{12.0}{\rmdefault}{\mddefault}{\updefault}{\color[rgb]{0,0,0}010}%
}}}
\put(751,689){\makebox(0,0)[lb]{\smash{\SetFigFont{10}{12.0}{\rmdefault}{\mddefault}{\updefault}{\color[rgb]{0,0,0}011}%
}}}
\put(1051,689){\makebox(0,0)[lb]{\smash{\SetFigFont{10}{12.0}{\rmdefault}{\mddefault}{\updefault}{\color[rgb]{0,0,0}100}%
}}}
\put(1351,689){\makebox(0,0)[lb]{\smash{\SetFigFont{10}{12.0}{\rmdefault}{\mddefault}{\updefault}{\color[rgb]{0,0,0}101}%
}}}
\put(1651,689){\makebox(0,0)[lb]{\smash{\SetFigFont{10}{12.0}{\rmdefault}{\mddefault}{\updefault}{\color[rgb]{0,0,0}110}%
}}}
\put(1951,689){\makebox(0,0)[lb]{\smash{\SetFigFont{10}{12.0}{\rmdefault}{\mddefault}{\updefault}{\color[rgb]{0,0,0}111}%
}}}
\put(-149,-961){\makebox(0,0)[lb]{\smash{\SetFigFont{10}{12.0}{\rmdefault}{\mddefault}{\updefault}{\color[rgb]{0,0,0}000}%
}}}
\put(151,-961){\makebox(0,0)[lb]{\smash{\SetFigFont{10}{12.0}{\rmdefault}{\mddefault}{\updefault}{\color[rgb]{0,0,0}001}%
}}}
\put(751,-961){\makebox(0,0)[lb]{\smash{\SetFigFont{10}{12.0}{\rmdefault}{\mddefault}{\updefault}{\color[rgb]{0,0,0}011}%
}}}
\put(1051,-961){\makebox(0,0)[lb]{\smash{\SetFigFont{10}{12.0}{\rmdefault}{\mddefault}{\updefault}{\color[rgb]{0,0,0}100}%
}}}
\put(1351,-961){\makebox(0,0)[lb]{\smash{\SetFigFont{10}{12.0}{\rmdefault}{\mddefault}{\updefault}{\color[rgb]{0,0,0}101}%
}}}
\put(1651,-961){\makebox(0,0)[lb]{\smash{\SetFigFont{10}{12.0}{\rmdefault}{\mddefault}{\updefault}{\color[rgb]{0,0,0}110}%
}}}
\put(151,-2611){\makebox(0,0)[lb]{\smash{\SetFigFont{10}{12.0}{\rmdefault}{\mddefault}{\updefault}{\color[rgb]{0,0,0}001}%
}}}
\put(-149,-2611){\makebox(0,0)[lb]{\smash{\SetFigFont{10}{12.0}{\rmdefault}{\mddefault}{\updefault}{\color[rgb]{0,0,0}000}%
}}}
\put(451,-961){\makebox(0,0)[lb]{\smash{\SetFigFont{10}{12.0}{\rmdefault}{\mddefault}{\updefault}{\color[rgb]{0,0,0}010}%
}}}
\put(451,-2611){\makebox(0,0)[lb]{\smash{\SetFigFont{10}{12.0}{\rmdefault}{\mddefault}{\updefault}{\color[rgb]{0,0,0}010}%
}}}
\put(751,-2611){\makebox(0,0)[lb]{\smash{\SetFigFont{10}{12.0}{\rmdefault}{\mddefault}{\updefault}{\color[rgb]{0,0,0}011}%
}}}
\put(1051,-2611){\makebox(0,0)[lb]{\smash{\SetFigFont{10}{12.0}{\rmdefault}{\mddefault}{\updefault}{\color[rgb]{0,0,0}100}%
}}}
\put(1351,-2611){\makebox(0,0)[lb]{\smash{\SetFigFont{10}{12.0}{\rmdefault}{\mddefault}{\updefault}{\color[rgb]{0,0,0}101}%
}}}
\put(1651,-2611){\makebox(0,0)[lb]{\smash{\SetFigFont{10}{12.0}{\rmdefault}{\mddefault}{\updefault}{\color[rgb]{0,0,0}110}%
}}}
\put(1951,-2611){\makebox(0,0)[lb]{\smash{\SetFigFont{10}{12.0}{\rmdefault}{\mddefault}{\updefault}{\color[rgb]{0,0,0}111}%
}}}
\put(2101,1439){\makebox(0,0)[lb]{\smash{\SetFigFont{10}{12.0}{\rmdefault}{\mddefault}{\updefault}{\color[rgb]{0,0,0}Match}%
}}}
\put(1986,-791){\makebox(0,0)[lb]{\smash{\SetFigFont{10}{12.0}{\rmdefault}{\mddefault}{\updefault}{\color[rgb]{0,0,0}111}%
}}}
\put(1136,124){\makebox(0,0)[lb]{\smash{\SetFigFont{10}{12.0}{\rmdefault}{\mddefault}{\updefault}{\color[rgb]{0,0,0}$U_f$}%
}}}
\put(1121,-1546){\makebox(0,0)[lb]{\smash{\SetFigFont{10}{12.0}{\rmdefault}{\mddefault}{\updefault}{\color[rgb]{0,0,0}$G$}%
}}}

{\color[rgb]{0,0,0}\put(-828,-436){\vector( 3,-1){675}}
}%
\put(-1349,-461){\makebox(0,0)[lb]{\smash{\SetFigFont{12}{14.4}{\rmdefault}{\mddefault}{\updefault}{\color[rgb]{0,0,0}Mean}%
}}}

\end{picture}

\end{center}
\caption{Mechanism of amplitude amplification for Grover's algorithm with $N=8$ and $M=1$.}
\label{GrovMech}
\end{figure}
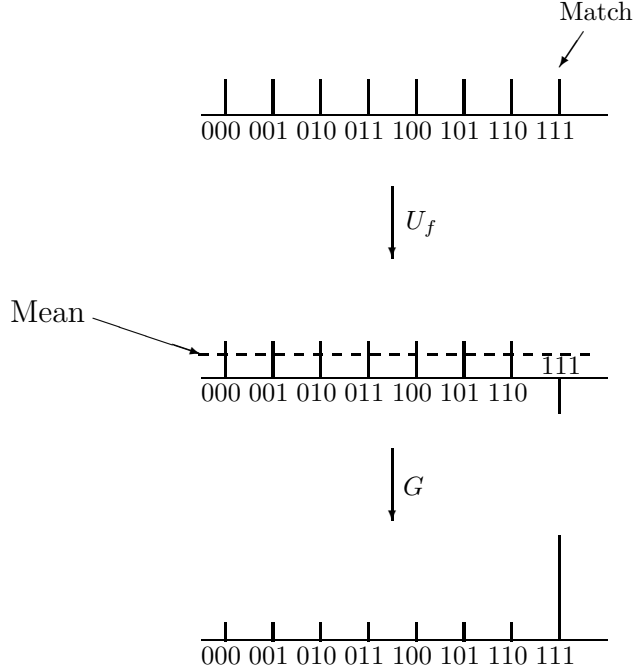
\end{center}

\begin{equation}
\label{SInteqn11}
\left| {W_3^{(G,1)} } \right\rangle = G\left| {W_2^{(G,1)} } \right\rangle =  
b_1^G \sum\limits_{i = 0}^{N - 1} {^{''} {\left| i \right\rangle} } 
+ a_1^G \sum\limits_{i = 0}^{N - 1}{^{'} {\left| i \right\rangle} },
\end{equation}

\noindent
such that,

\begin{equation}
\label{SInteqn12}
M(a_1^G)^2+(N - M)(b_1^G)^2   = 1.
\end{equation}

\item[5-] {\it Measurement}. Measure the first $n$ qubits. The probabilities of the system will be as follows:

\begin{itemize}
\item[i-] Probability $P_s^{\left( 1 \right)_G }$ to find a match out of the $M$ possible matches can be calculated as follows:

\begin{equation}
\label{SInteqn13}
\begin{array}{l}
P_s^{\left( 1 \right)_G } = M(a_1^G)^2\\
\,\,\,\,\,\,\,\,\,\,\,\,\,\,\,  = 9\left( {\frac{M}{N}} \right) - 24\left( {\frac{M}{N}} \right)^2  + 16\left( {\frac{M}{N}} \right)^3. \\ 
\end{array}
\end{equation}

\item[ii-] Probability $P_{ns}^{\left( 1 \right)_G }$ to find undesired result out of the states can be 
calculated as follows:

\begin{equation}
\label{SInteqn14}
P_{ns}^{\left( 1 \right)_G } = (N - M)(b_1^G)^2. 
\end{equation}
\end{itemize}

Notice that, using Eqn. \ref{SInteqn12},

\begin{equation}
\label{SInteqn15}
 P_s^{\left( 1 \right)_G } + P_{ns}^{\left( 1 \right)_G } = M(a_1^G)^2 + (N - M)(b_1^G)^2  = 1. \\ 
\end{equation}
\end{itemize}



\subsubsection*{Performance of a Single Iteration}
\label{Goneitr}

\begin{table}[t]
\begin{center}
\begin{tabular}
{|c|c|c|c|}
\hline
 $n$, where $N=2^n$ & 
\ Max. prob.  & 
 Min. prob.  & 
 Avg. prob.   \\ \hline
2  & 
1.0  & 
0.0  & 
0.5  \\ \hline
 3  & 
 1.0  & 
 0.0  & 
 0.5  \\ \hline
 4  & 
 1.0  & 
 0.0  & 
0.5 \\ \hline
 5  & 
 1.0  & 
 0.0  & 
 0.5 \\ \hline
 6  & 
 1.0  & 
 0.0  & 
 0.5  \\ \hline

\end{tabular}
\caption {Performance of the first iteration of Grover's algorithm with different size search space.}
\label{SInttab1}
\end{center}
\end{table}

\begin{figure}[t]

\centerline{\includegraphics[width=4.0in,height=3.0in]{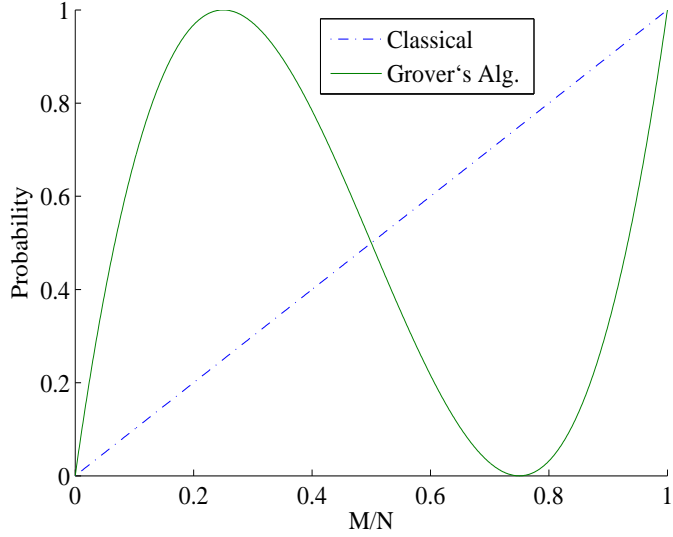}}
\caption{Probabilities of success $P_s^{\left( 1 \right)_G }$ and $P_s^{(classical)}$ as a function of ${M}/{N}$.} 
\label{SIntfig3}
\end{figure}

Considering Eqn. \ref{SInteqn9}, Eqn. \ref{SInteqn10}, Eqn. \ref{SInteqn11} and Eqn. \ref{SInteqn13}, we can see that 
the probability to find a solution for a fixed size search space varies according to the number of matches $M$ 
in the superposition (see Fig. \ref{SIntfig3}). 

From Tab. \ref{SInttab1}, we can see that the maximum probability of success is always 1.0,
and the minimum probability (worst case) is always 0.0. The average 
probability over the possible oracles for different number of matches is always 0.5. 
It implies that the average performance of the first iteration remains constant even 
with the increase of the size of the list.

To verify these results, taking into account that the oracle $U_f$ is taken as a black box, we can define 
the average probability, $average(P_s^{\left( 1 \right)_G })$, as follows:


\begin{equation}
\label{SInteqn16}
\begin{array}{l}
 average\left( P_s^{\left( 1 \right)_G } \right) = \frac{1}{{2^N }}\sum\limits_{M = 1}^N {{}^NC_M } P_s^{\left( 1 \right)_G }  \\ 
  \,\,\,\,\,\,= \frac{1}{{2^N }}\sum\limits_{M = 1}^N {\frac{{N!}}{{M!\left( {N - M} \right)!}}} \left( {9\frac{M}{N} - 24\left( {\frac{M}{N}} \right)^2  + 16\left( {\frac{M}{N}} \right)^3 } \right) \\ 
  \,\,\,\,\,\,= \frac{1}{{2^N }}\left( {\sum\limits_{M = 1}^N {\frac{{9\left( {N - 1} \right)!}}{{\left( {M - 1} \right)!\left( {N - M} \right)!}}}  - \sum\limits_{M = 1}^N {\frac{{24\,M\,\left( {N - 1} \right)!}}{{N\,\left( {M - 1} \right)!\left( {N - M} \right)!}}}  + \sum\limits_{M = 1}^N {\frac{{16\,\,M^2 \,\,\left( {N - 1} \right)!}}{{N^2 \,\,\left( {M - 1} \right)!\left( {N - M} \right)!}}} } \right) \\ 
  \,\,\,\,\,\,= \frac{1}{{2^N }}\left( {9.\,2^{N - 1}  - 24\left(\frac{{2^{N - 1}  + \left( {N - 1} \right)2^{N - 2} }}{N}\right) + 16\left(\frac{{ 3N2^N + N^2 2^N }}{{8\,N^2 }}\right)} \right) \\ 
  \,\,\,\,\,\,= \frac{1}{{2^N }}\left( {9.\,2^{N - 1}  - 6.\,2^N \left(\frac{{N + 1}}{N}\right) + 2^N \left(\frac{{6N + 2N^2 }}{{N^2 }}\right)} \right) \\ 
  \,\,\,\,\,\,= \frac{{2^{N - 1} }}{{2^N }}\left( {9\, - 12\left(\frac{{N + 1}}{N}\right) + \frac{{12N + 4N^2 }}{{N^2 }}} \right) \\ 
  \,\,\,\,\,\,= \frac{1}{2}, \\ 
 \end{array}
\end{equation}

\noindent
where ${}^NC_M  = \frac{{N!}}{{M!(N - M)!}}$ is the number of possible cases for $M$ matches. We can see that as the 
size of the list increases $(N \to \infty)$, $average (P_s^{\left( 1 \right)_G })$ shown in 
Eqn. \ref{SInteqn16} remains one-half.

Classically, we can do a single trial guess to find any match. We may succeed in finding a solution with probability 
$P^{(classical)}_{s} = {M}/{N}$. The average probability can be calculated as follows:

\begin{equation}
\label{SInteqn17}
\begin{array}{l}
average(P_s^{(classical)} ) = \frac{1}{{2^N }}\sum\limits_{M = 1}^N {{}^NC_M P_s^{(classical)} } \\
\,\,\,\,\,\,\,\,\,\,\,\,\,\,\,\,\,\,\,\,\,\,\,\,\,\,\,\,\,\,\,\,\,\,\,\,\,\,\,\,\,\,\,\,\,\,\,\,\,\,\,= \frac{1}{{2^N }}\sum\limits_{M = 1}^N {\frac{{(N-1)!}}{{(M-1)!(N - M)!}}} \\
\,\,\,\,\,\,\,\,\,\,\,\,\,\,\,\,\,\,\,\,\,\,\,\,\,\,\,\,\,\,\,\,\,\,\,\,\,\,\,\,\,\,\,\,\,\,\,\,\,\,\,= \frac{1}{2}. \\ 
\end{array}
\end{equation}

It means that we have an average probability one-half to find or not to find a solution by a single random guess, even 
with the increase in the number of matches, similar to the first iteration of Grover's algorithm.

To compare the performance of the first iteration of Grover's algorithm and the classical guess 
technique, Fig. \ref{SIntfig3} shows the probability of success of the two algorithms just mentioned as a function of $0<{M}/{N}\le1$.


We can see from Fig. \ref{SIntfig3} that Grover's algorithm solves the case where $M={N}/{4}$ with certainty. 
The probability of success of Grover's algorithm will be below one-half for $M>{N}/{2}$ and will fail with 
certainty for $M={3N}/{4}$. The probability of success of the classical guess technique is always over that of Grover's 
algorithm for $M>{N}/{2}$.

\subsubsection{Iterating the Algorithm}

Before we go further in the analysis of the algorithm after arbitrary number of iterations $q_G$, 
we will re-formulate the equations of the first iteration according to the way used in \cite{boyer96}.

Initially before the first iteration, we had $M$ states with amplitude 
$a_0^G= {1}/{\sqrt{N}}$ and $N-M$ states with amplitude $b_0^G= {1}/{\sqrt{N}}$. 
After applying the oracle $U_f$ and the diffusion operator $G$, the new amplitudes $a_1^G$ and $b_1^G$ 
can be re-written as follows:

\begin{equation}
\label{SInteqn18}
a_1^G  = \frac{{N - 2M}}{N}a_0^G  + \frac{{2\left( {N - M} \right)}}{N}b_0^G ,\,\,\,b_1^G  = \frac{{N - 2M}}{N}b_0^G  - \frac{{2M}}{N}a_0^G. 
\end{equation}

\pagebreak
The iterative version of the algorithm can be summarised as follows:

\begin{itemize}

\item[1-]Prepare a quantum register of $n+1$ 
qubits. The first $n$ qubits all in state $\left| 0 \right\rangle$ and the extra qubit in state 
$\left| 1 \right\rangle$.

\item[2-]Apply the Hadamard gate on each of the $n+1$ qubits in 
parallel. 
 
\item[3-]Iterate the following steps $q_G$ times,
\begin{itemize}
 
\item[i-]Apply the oracle $U_{f}$.

\item[ii-]Apply the diffusion operator $G$ on the first $n$ qubits.
\end{itemize}

\item[4-] Measure the first $n$ qubits to get the result with probability $P^{(q_G)}_{s}$.
\end{itemize}

The system after $q_G\ge1$ iterations can be written as follows,

\begin{equation}
\label{SInteqn11q}
\left| {W^{(G,q_G)} } \right\rangle = b_q^G \sum\limits_{i = 0}^{N - 1} {^{''} {\left| i \right\rangle} } 
+ a_q^G \sum\limits_{i = 0}^{N - 1}{^{'} {\left| i \right\rangle} },
\end{equation}

\noindent
such that,

\begin{equation}
\label{SInteqn12q}
M(a_q^G)^2+(N - M)(b_q^G)^2   = 1,
\end{equation}

\noindent
where the amplitudes $a_q^G$ and $b_q^G$ after $q_G\ge1$ iterations are defined by the following recurrence 
relations \cite{boyer96},

\begin{equation}
\begin{array}{l}
a_0^G=b_0^G=\frac{1}{\sqrt{N}},\\
a_q^G  = \frac{{N - 2M}}{N}a_{q-1}^G  + \frac{{2\left( {N - M} \right)}}{N}b_{q-1}^G ,\,\,\,b_q^G  = \frac{{N - 2M}}{N}b_{q-1}^G  - \frac{{2M}}{N}a_{q-1}^G.\\
\end{array} 
\end{equation}

Solving these recurrence relations, the closed forms can be written as follows \cite{boyer96}:

\begin{equation}
\label{SInteqn19}
a_q^G  = \frac{1}{{\sqrt M }}\sin \left( {\left( {2q_G  + 1} \right)\theta _G } \right),\,\,\,b_q^G  = \frac{1}{{\sqrt {N - M} }}\cos \left( {\left( {2q_G  + 1} \right)\theta _G } \right),
\end{equation}

\noindent
where $\sin ^2 \left( \theta _G\right)  = {M}/{N}$ and $0 < \theta _G  \le {\pi}/{2}$. 

\pagebreak
The probabilities of the system will be 
as follows:
\begin{itemize}
\item[1-] The probability of success after $q_G$ iterations is:

\begin{equation}
\label{SInteqn20}
 P^{(q_G)}_{s} =M \left(a_q^G\right)^2 = \sin ^2 ((2q_G + 1)\theta_G ).
\end{equation}

\item[2-] The probability of failure after $q_G$ iterations is:

\begin{equation}
\label{SInteqn20b}
 P^{(q_G)}_{ns} =\left( N-M \right) \left(b_q^G\right)^2 = \cos ^2 ((2q_G + 1)\theta_G ).
\end{equation}
\end{itemize}



The aim is to find a solution with probability as close as possible to certainty. 
It was shown in \cite{boyer96} that $P^{(\overline q_G)}_{ns}=0$ when 
$\overline q _G  = \left( {\pi  - 2\theta _G } \right)/4\theta _G $, but since the number of iterations must be 
integer, let $q_G  = \left\lfloor \pi /4\theta _G \right\rfloor$ where $\left| {q_G  - \overline q _G } \right| \le 1/2$. 
And since, $\sin ^2 \left( \theta_G  \right) = {M}/{N}$, we have for small $M/N$ that, $\theta_G  \ge \sin \left( \theta_G  \right) 
= \sqrt {{M}/{N}} $, then,

\begin{equation}
\label{SInteqn21}
q_G  = \left\lfloor {\frac{\pi }{{4\theta _G }}} \right\rfloor \le  {\frac{\pi }{{4\theta _G }}} \le \frac{\pi }{4}\sqrt {\frac{N}{M}}  = O\left( {\sqrt {\frac{N}{M}} } \right),
\end{equation}
where $ \left\lfloor {\,\,} \right\rfloor$ is the floor operation. 
The lower bound of the probability of success using $q_G$ is $P^{(q_G)}_{s} \ge 1 - {M}/{N}\ge 0$, 
which is negligible only for small $M/N$.

\begin{figure}[htbp]
\centerline{\includegraphics[width=4.00in,height=3.0in]{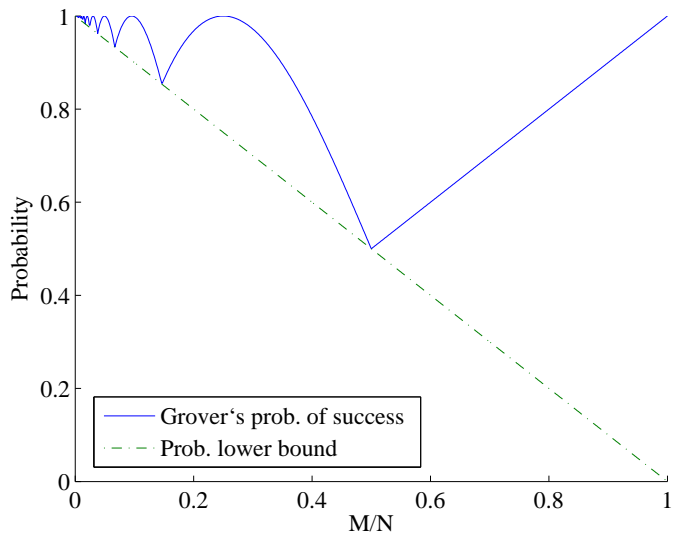}}
\caption{Probability of success of Grover's algorithm using the required number of iterations $q_G$.} 
\label{SIntfig4}
\end{figure}

To demonstrate the real behaviour of Grover's algorithm, we may plot the 
probability of success $P^{(q_G)}_{s}$ using the required number of iterations $q_G$ for any given $M$. 
Fig. \ref{SIntfig4} shows this behaviour as a function of $0<{M}/{N}\le1$. We can see from the plot that the minimum probability that Grover's algorithm may reach is approximately 50.0{\%} 
when ${M}/{N} = 0.5$. The algorithm will behave similar to the classical single random guess for ${M}/{N} > 0.5$ 
since $q_G=0$ in that range. For $0.145<M/N\le0.5$, $q_G=1$ where we can see that for ${M}/{N} = 0.25$ the algorithm 
will succeed with certainty after a single iteration. For $M/N<0.145$, $q_G>1$ where the algorithm will behave more reliably. 
In an attempt to avoid this drawback in the behaviour for multiple matches, it was proposed in \cite{nc00a} that we can 
double the search space by adding $N$ non-match items so that the number of matches will always be less than half the search space 
and iterate the algorithm $\pi /4\sqrt {2N/M}$ instead of $\pi /4\sqrt {N/M}$ so it still runs in 
$O\left( {\sqrt {N/M} } \right)$. Using this approach will increase the space/time requirements to still get the result with 
probability at least one-half when $M=N$, where we can get the result with certainty in this case if we did not use that approach.

\subsection{Number of Matches is Unknown}
\label{unknowG}
In case the number of matches $M$ is unknown, an algorithm that employs Grover's 
algorithm \cite{boyer96} can be used for $1 \le M\le {3N}/{4}$ which can be summarised as follows:

\begin{itemize}
\item[1-] Start with $m=1$ and $\lambda= {8}/{7}$. $(\lambda$ can take any value between 
1 and ${4}/{3})$
\item[2-]Pick an integer $j$ between 0 and $m-1$ in a uniform random manner.
\item[3-]Run $j$ iterations of Grover's algorithm on the state: 
$\frac{1}{{\sqrt N }}\sum\limits_{i = 0}^{N - 1} {\left| i \right\rangle }$.

\item[4-]Measure the register and assume $i$ is the output.
\item[5-]If $f(i)=1$, then we found a solution and exit.
\item[6-]Let $m=min\left( \lambda m,\sqrt{N}\right)$ and go to step 2.
\end{itemize}

\begin{figure}[H]
\centerline{\includegraphics[width=4.00in,height=3.0in]{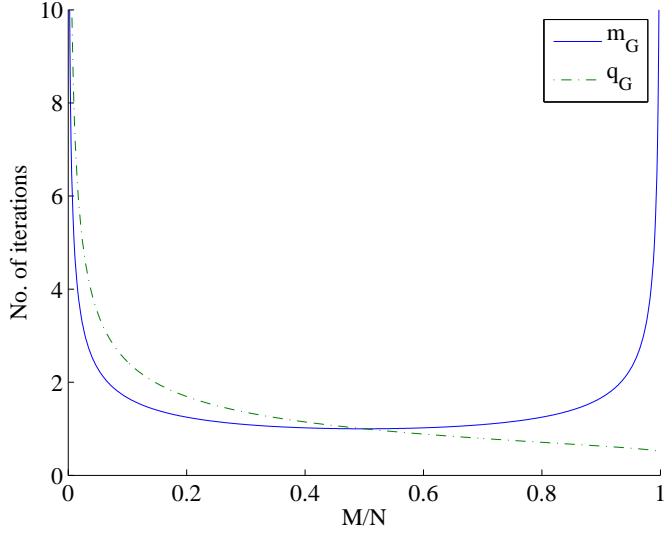}}
\caption{The actual behaviour of the functions representing the required number of iterations for known ($q_G=\pi /4\theta _G$) 
and unknown ($m_G=1 /\sin \left( {2\theta _G } \right)$) number of solutions, where the number of iterations is the flooring of the values (step function). } 
\label{SIntfig5}
\end{figure}

It was shown that the total expected number of iterations is approximately $8m_G \approx 4\sqrt {N/M}$ for small $M/N$ , 
where $m_G  \ge 1/\sin \left( {2\theta _G } \right)= O\left( {\sqrt {N/M} } \right)$ for $M \le {3N}/{4}$. 
The algorithm works only for $1 \le M\le {3N}/{4}$, where for $M> {3N}/{4}$, a classical sampling techniques can be used. 

The reason that this algorithm will fail for $M> {3N}/{4}$ is that $m_G$ is acting as a lower bound for $q_G$ for 
$M\le N/2$. It handles the case where $q_G=0$ in a constant manner for 
$N/2 <M\le3N/4$. However, it will increase exponentially for $M> {3N}/{4}$ where it is no longer able to approximate $q_G$, i.e. using 
the algorithm in that range means that the expected number of iterations will increase exponentially where the problem 
should be easier, as shown in Fig. \ref{SIntfig5}.

\section{Conclusion}

In this paper, we analysed Grover's algorithm over the whole search space. We found that, although Grover's 
algorithm is optimal \cite{Zalka99} for a single match in the search space, its reliability may 
decrease for multiple matches, i.e. the behaviour of the algorithm is not reliable 
over the whole range where the minimum probability it may reach is approximately 50.0{\%} when ${M}/{N}= 0.5$. 
The best behaviour is for $M/N<0.145$ and in the neighbourhood of $M/N=0.25$. 
The role of Grover's algorithm disappears for ${M}/{N} > 0.5$ where the required number of iterations will 
vanish in that range. Grover's algorithm may not be suitable for 
practical implementation since a practical quantum algorithm should be able to handle both the easiest cases 
and the hardest cases.



\bibliography{swgrover}
\bibliographystyle{plain}

\end{document}